\documentclass[aps, prd, twocolumn, superscriptaddress, showpacs, amsmath, amssymb, tightenlines]{revtex4-1}

\usepackage{graphicx} 
\usepackage{dcolumn}  
\usepackage{epstopdf}
\usepackage{bm}

\usepackage[absolute]{textpos}

\graphicspath{{ps}}

\newcommand{\bsig}{B_{\textrm{sig}}}
\newcommand{\btag}{B_{\textrm{tag}}}
\newcommand{\ecl}{E_{\textrm{ECL}}}
\newcommand{\psig}{p_{\textrm{sig}}^*}

\newcommand{\bdl}{\cos \theta_{B,D^{(*)}\ell}}
\newcommand{\tautoe}{\tau^+ \to e^+ \bar{\nu}_\tau \nu_e}
\newcommand{\tautomu}{\tau^+ \to \mu^+ \bar{\nu}_\tau \nu_\mu}
\newcommand{\tautopi}{\tau^+ \to \pi^+ \bar{\nu}_\tau}
\newcommand{\tautorho}{\tau^+ \to \rho^+ \bar{\nu}_\tau}
\newcommand{\Bdecay}{B^+ \to \tau^+ \nu_\tau}
\newcommand{\bbar}{B \bar{B}}

\newcommand{\mevcc}{{\mathrm{\,Me\kern -0.1em V\!/}c^2}}
\newcommand{\mevc}{{\mathrm{\,Me\kern -0.1em V\!/}c}}
\newcommand{\mev}{{\mathrm{\,Me\kern -0.1em V\!}}}
\newcommand{\gevcc}{{\mathrm{\,Ge\kern -0.1em V\!/}c^2}}
\newcommand{\gevc}{{\mathrm{\,Ge\kern -0.1em V\!/}c}}
\newcommand{\gev}{{\mathrm{\,Ge\kern -0.1em V\!}}}

\begin{document}

%
%


\title{\boldmath Measurement of the branching fraction of $\Bdecay$ decays with the semileptonic tagging method}


\noaffiliation
\affiliation{University of the Basque Country UPV/EHU, 48080 Bilbao}
\affiliation{University of Bonn, 53115 Bonn}
\affiliation{Novosibirsk State University, Novosibirsk 630090}
\affiliation{Faculty of Mathematics and Physics, Charles University, 121 16 Prague}
\affiliation{Chonnam National University, Kwangju 660-701}
\affiliation{University of Cincinnati, Cincinnati, Ohio 45221}
\affiliation{Deutsches Elektronen--Synchrotron, 22607 Hamburg}
\affiliation{Justus-Liebig-Universit\"at Gie\ss{}en, 35392 Gie\ss{}en}
\affiliation{The Graduate University for Advanced Studies, Hayama 240-0193}
\affiliation{Hanyang University, Seoul 133-791}
\affiliation{University of Hawaii, Honolulu, Hawaii 96822}
\affiliation{High Energy Accelerator Research Organization (KEK), Tsukuba 305-0801}
\affiliation{IKERBASQUE, Basque Foundation for Science, 48013 Bilbao}
\affiliation{Indian Institute of Technology Guwahati, Assam 781039}
\affiliation{Indian Institute of Technology Madras, Chennai 600036}
\affiliation{Indiana University, Bloomington, Indiana 47408}
\affiliation{Institute of High Energy Physics, Chinese Academy of Sciences, Beijing 100049}
\affiliation{Institute of High Energy Physics, Vienna 1050}
\affiliation{Institute for High Energy Physics, Protvino 142281}
\affiliation{INFN - Sezione di Torino, 10125 Torino}
\affiliation{Institute for Theoretical and Experimental Physics, Moscow 117218}
\affiliation{J. Stefan Institute, 1000 Ljubljana}
\affiliation{Kanagawa University, Yokohama 221-8686}
\affiliation{Institut f\"ur Experimentelle Kernphysik, Karlsruher Institut f\"ur Technologie, 76131 Karlsruhe}
\affiliation{Department of Physics, Faculty of Science, King Abdulaziz University, Jeddah 21589}
\affiliation{Korea Institute of Science and Technology Information, Daejeon 305-806}
\affiliation{Korea University, Seoul 136-713}
\affiliation{Kyungpook National University, Daegu 702-701}
\affiliation{\'Ecole Polytechnique F\'ed\'erale de Lausanne (EPFL), Lausanne 1015}
\affiliation{Faculty of Mathematics and Physics, University of Ljubljana, 1000 Ljubljana}
\affiliation{Ludwig Maximilians University, 80539 Munich}
\affiliation{Luther College, Decorah, Iowa 52101}
\affiliation{University of Maribor, 2000 Maribor}
\affiliation{Max-Planck-Institut f\"ur Physik, 80805 M\"unchen}
\affiliation{School of Physics, University of Melbourne, Victoria 3010}
\affiliation{Moscow Physical Engineering Institute, Moscow 115409}
\affiliation{Moscow Institute of Physics and Technology, Moscow Region 141700}
\affiliation{Graduate School of Science, Nagoya University, Nagoya 464-8602}
\affiliation{Kobayashi-Maskawa Institute, Nagoya University, Nagoya 464-8602}
\affiliation{Nara Women's University, Nara 630-8506}
\affiliation{National Central University, Chung-li 32054}
\affiliation{National United University, Miao Li 36003}
\affiliation{Department of Physics, National Taiwan University, Taipei 10617}
\affiliation{H. Niewodniczanski Institute of Nuclear Physics, Krakow 31-342}
\affiliation{Niigata University, Niigata 950-2181}
\affiliation{University of Nova Gorica, 5000 Nova Gorica}
\affiliation{Novosibirsk State University, Novosibirsk 630090}
\affiliation{Osaka City University, Osaka 558-8585}
\affiliation{Pacific Northwest National Laboratory, Richland, Washington 99352}
\affiliation{Peking University, Beijing 100871}
\affiliation{University of Pittsburgh, Pittsburgh, Pennsylvania 15260}
\affiliation{University of Science and Technology of China, Hefei 230026}
\affiliation{Seoul National University, Seoul 151-742}
\affiliation{Soongsil University, Seoul 156-743}
\affiliation{Sungkyunkwan University, Suwon 440-746}
\affiliation{School of Physics, University of Sydney, NSW 2006}
\affiliation{Department of Physics, Faculty of Science, University of Tabuk, Tabuk 71451}
\affiliation{Tata Institute of Fundamental Research, Mumbai 400005}
\affiliation{Excellence Cluster Universe, Technische Universit\"at M\"unchen, 85748 Garching}
\affiliation{Tohoku University, Sendai 980-8578}
\affiliation{Department of Physics, University of Tokyo, Tokyo 113-0033}
\affiliation{Tokyo Institute of Technology, Tokyo 152-8550}
\affiliation{Tokyo Metropolitan University, Tokyo 192-0397}
\affiliation{University of Torino, 10124 Torino}
\affiliation{Utkal University, Bhubaneswar 751004}
\affiliation{CNP, Virginia Polytechnic Institute and State University, Blacksburg, Virginia 24061}
\affiliation{Wayne State University, Detroit, Michigan 48202}
\affiliation{Yamagata University, Yamagata 990-8560}
\affiliation{Yonsei University, Seoul 120-749}
  \author{B.~Kronenbitter}\affiliation{Institut f\"ur Experimentelle Kernphysik, Karlsruher Institut f\"ur Technologie, 76131 Karlsruhe} 
  \author{M.~Heck}\affiliation{Institut f\"ur Experimentelle Kernphysik, Karlsruher Institut f\"ur Technologie, 76131 Karlsruhe} 
  \author{P.~Goldenzweig}\affiliation{Institut f\"ur Experimentelle Kernphysik, Karlsruher Institut f\"ur Technologie, 76131 Karlsruhe} 
  \author{T.~Kuhr}\affiliation{Ludwig Maximilians University, 80539 Munich} 
  \author{A.~Abdesselam}\affiliation{Department of Physics, Faculty of Science, University of Tabuk, Tabuk 71451} 
  \author{I.~Adachi}\affiliation{High Energy Accelerator Research Organization (KEK), Tsukuba 305-0801}\affiliation{The Graduate University for Advanced Studies, Hayama 240-0193} 
  \author{H.~Aihara}\affiliation{Department of Physics, University of Tokyo, Tokyo 113-0033} 
  \author{S.~Al~Said}\affiliation{Department of Physics, Faculty of Science, University of Tabuk, Tabuk 71451}\affiliation{Department of Physics, Faculty of Science, King Abdulaziz University, Jeddah 21589} 
  \author{K.~Arinstein}\affiliation{Budker Institute of Nuclear Physics SB RAS, Novosibirsk 630090}\affiliation{Novosibirsk State University, Novosibirsk 630090} 
  \author{D.~M.~Asner}\affiliation{Pacific Northwest National Laboratory, Richland, Washington 99352} 
% ...
\collaboration{The Belle Collaboration}

\begin{abstract}
We report a measurement of the branching fraction of $\Bdecay$ decays using a data sample of $772 \times 10^6 \bbar$ pairs, collected at the $\Upsilon(4S)$ resonance with the Belle detector at the KEKB asymmetric-energy $e^+e^-$ collider. We reconstruct the accompanying $B$ meson in a semileptonic decay and detect the recoiling $B$ candidate in the decay channel $\Bdecay$.
We obtain a branching fraction of ${\cal B}(\Bdecay) = [1.25 \pm 0.28 ({\rm stat.}) \pm 0.27({\rm syst.})] \times 10^{-4}$. This result is in good agreement with previous measurements and the expectation from the calculation based on the Standard Model.
\end{abstract}

\pacs{13.20.He, 14.40.Nd}

\maketitle


{\renewcommand{\thefootnote}{\fnsymbol{footnote}}}
\setcounter{footnote}{0}
In the Standard Model (SM), the branching fraction of the purely leptonic decay $\Bdecay$~\cite{CC} is given by
\begin{equation}
 {\cal B}(\Bdecay)_{\rm SM} = \frac{G_F^2 m_B m_\tau^2}{8 \pi} \left( 1 - \frac{m_\tau^2}{m_B^2} \right)^2 f_B^2 |V_{ub}|^2 \tau_B, \label{equ:BR}
\end{equation}
where $G_F$ is the Fermi coupling constant, $V_{ub}$ the Cabibbo-Kobayashi-Maskawa matrix element, $m_B$ and $m_\tau$ the masses of the $B$ meson and the $\tau$ lepton, respectively, $\tau_B$ the lifetime of the $B$ meson, and $f_B$ the $B$-meson decay constant.
The branching fraction depends strongly on the mass of the lepton due to helicity suppression, and thus $\Bdecay$ is expected to have the largest leptonic branching fraction of the $B^+$ meson and is the only decay of this kind for which there is experimental evidence.
All of the inputs of Eq.~(\ref{equ:BR}) are measured experimentally~\cite{PDG} except for $f_B$, which is calculated in the framework of lattice quantum chromodynamics~\cite{LQCD}. 
An estimate of the branching fraction, which uses a unitarity-constrained $V_{ub}$ value aggregated from other measurements and lattice calculations of $f_B$, gives ${\cal B}(\Bdecay) = (0.75^{+0.10}_{-0.05}) \times 10^{-4}$~\cite{CKMfitter}.

\par
Physics beyond the SM, such as the presence of additional charged Higgs bosons~\cite{NewPhys1,NewPhys2}, could interfere constructively or destructively with the SM weak decay process. Measurements by the BaBar~\cite{BaBarLep,BaBarHad} and Belle~\cite{BelleLep} collaborations showed a slight disagreement with the SM expectation, but the most recent measurement by Belle~\cite{BelleHad}, using a hadronic tagging method, is in very good agreement. The current world average, ${\cal B}(\Bdecay) = (1.14 \pm 0.27) \times 10^{-4}$~\cite{PDG}, shows no sign of physics beyond the SM. 

\par
The analysis described here contains the following improvements over our previous measurement~\cite{BelleLep}: an improved semileptonic tagging method; the reconstruction of an additional $\tau$ decay channel; a newly optimized selection, including a dedicated suppression of background containing converted photons and a multivariate suppression of continuum $e^+ e^- \to q \bar{q} \ (q = u, d, s, c)$ background; the inclusion of a second variable, the momentum of the decay product of the $\tau$, in the final fit; and the usage of the full Belle data set, which contains almost 20\% more data than in the previous analysis.

\par
The measurement is performed using the final Belle data sample consisting of an integrated luminosity of $711 \,{\rm fb}^{-1}$, containing $(772 \pm 11) \times 10^6 \bbar$ pairs, collected at the $\Upsilon(4S)$ resonance at the KEKB asymmetric-energy $e^+e^-$ collider~\cite{kekb}.
We use a smaller data sample with an integrated luminosity of $79 \,{\rm fb}^{-1}$ taken at an energy $60 \mev$ below the $\Upsilon(4S)$ mass to study background from continuum $e^+ e^- \to q \bar{q}$ events.
We generate Monte Carlo (MC) samples of signal and background events. We model the decays of unstable particles using the software package EvtGen~\cite{evtgen} and we simulate the detector response using GEANT3~\cite{GEANT}. The simulated signal events are overlaid with beam-related background events that were recorded with a random trigger.

\par 
The Belle detector is a large-solid-angle magnetic spectrometer that consists of a silicon vertex detector (SVD), a 50-layer central drift chamber (CDC), an array of aerogel threshold Cherenkov counters (ACC), a barrel-like arrangement of time-of-flight scintillation counters (TOF), and an electromagnetic calorimeter (ECL) composed of CsI(Tl) crystals located inside a superconducting solenoid coil that provides a 1.5~T magnetic field. An iron flux-return located outside the coil (KLM) is instrumented to detect $K_L^0$ mesons and to identify muons. The detector is described in detail elsewhere~\cite{Belle}.
Two SVD configurations were used. A 2.0 cm beam pipe and a three-layer SVD were used for the first sample of $152 \times 10^6 B\bar{B}$ pairs, while a 1.5 cm beam pipe, a four-layer SVD and a small-cell inner drift chamber were used to record the remaining $620 \times 10^6 B\bar{B}$ pairs~\cite{svd2}.

\par 
Since the detectable signature of a $\Bdecay$ decay is often only a single charged track and thus inadequate for an efficient signal-background discrimination, we reconstruct the accompanying $B$ meson (referred to as $\btag$) in the semileptonic decay channels $B^+ \to \bar{D}^{*0} \ell^+ \nu_\ell$ and $\bar{D}^{0} \ell^+ \nu_\ell$, where $\ell$ can be an electron or muon. The $D^{*0}$ mesons are reconstructed through the decays $D^{*0} \to D^0 \pi^0$ and $D^0 \gamma$ and the $D^0$ mesons through the decays $D^0 \to K^- \pi^+ \pi^0$, $K^- \pi^+ \pi^+ \pi^-$, $K^0_S \pi^+ \pi^- \pi^0$, $K^- \pi^+$, $K^0_S \pi^+ \pi^-$, $\pi^+ \pi^- \pi^0$, $K^0_S \pi^0$, $K^0_S K^+ K^-$, $K^+ K^-$, and $\pi^+ \pi^-$. Neutral pions are reconstructed as $\pi^0 \to \gamma \gamma$ and $K_S^0$ as $K_S^0 \to \pi^+ \pi^-$.

\par
To maximize the efficiency in identifying $\btag$ candidates, loose selection criteria are applied throughout their reconstruction. Charged final state particles are selected from well-measured tracks and are required to have a distance to the interaction point along (perpendicular to) the beam direction, denoted as $dz~(dr)$, of less than $4~(2)$~cm. Photons used for the reconstruction of neutral pions are required to have an energy of at least $30 \mev$ and the invariant mass of the two-photon system ($M_{\gamma \gamma}$) must satisfy $|M_{\gamma \gamma} - m_{\pi^0}| < 19~\mevcc$; this corresponds to a width of $3.2$ standard deviations ($\sigma$). The invariant mass of the two charged tracks that are used to form $K_S^0$ candidates must lie within $30~\mevcc~(4.5 \sigma)$ of the nominal $K_S^0$ mass. The momenta of $D^{(*)0}$ meson candidates are required to be below $2.5 \gevc$ to reject $D^{(*)0}$ mesons from $e^+e^- \to c \bar{c}$ events.

All further selections related to the $\btag$ candidate are performed by training multiple instances of a multivariate selection (MVS) method based on the NeuroBayes package~\cite{NB}. An MVS classifier is trained for each of the reconstructed decay channels, where a large sample of simulated $B$ mesons that decay generically is used. The variables used in the training of the MVS related to the intermediate particles are identical to those used for the hadronic full-reconstruction method~\cite{HadFR} and the same hierarchical approach is used. The mass, momentum, and decay channel of the particle candidate, as well as the momenta, angles, and the output of the separately-trained MVS of daughter particles are used in the training, if applicable, in addition to particle-specific information like the output of the particle identification. The training related to the $\btag$ candidate is performed using the following information, sorted by their discriminating power in descending order: the outputs of the MVS of the decay products; the mass of the $D^0$ meson candidate or the difference of the masses of the $D^{*0}$ and the $D^0$ meson candidates, depending on the decay channel; the angle between the $D^{(*)0}$ meson candidate and the lepton in the center-of-mass system of the $\Upsilon(4S)$ (CM); the angle between the $\btag$ candidate and the $D^{(*)0}$ meson candidate in the center-of-mass system of the $\btag$ candidate; the distance of minimum approach between the $D^{(*)0}$ decay vertex and the trajectory of the lepton; the decay channel of the $D^{(*)0}$ meson; and the angle of the reconstructed $\btag$ candidate with respect to the beam axis in the CM.

\par
The training variables were chosen to be uncorrelated with the cosine of the angle between the momentum of the $B$ meson and the $D^{(*)}\ell$ system, calculated under the assumption that only one massless particle is not reconstructed. It is given by 
\begin{equation}
\bdl = \frac{2 E_{\rm beam} E_{D^{(*)} \ell} - m_B^2 c^4 - m^2_{D^{(*)}\ell} c^4}{2 p^*_B p^*_{D^{(*)} \ell} c^2}, \label{equ:bdl}
\end{equation}
where $p^*_B$ is the nominal $B$ meson momentum calculated from the beam energy and the nominal mass, $E_{D^{(*)} \ell}$, $m_{D^{(*)} \ell}$, and $p^*_{D^{(*)} \ell}$ are the energy, mass, and momentum of the $D^{(*)} \ell$ system, respectively, and $E_{\rm beam}$ is the energy of the beam (all in the CM). 
While the independence between the angle and the output of the MVS is not exploited in this analysis, it can be used in principle to produce signal-free sideband samples to study backgrounds and to extract the number of correctly reconstructed $\btag$ candidates. 
It is also used later for a refined selection since correctly reconstructed $\btag$ candidates have values between $-1$ and $1$, while background events where the assumption of only one missing massless particle fails have a much larger range of values. Some $\btag$ candidates are reconstructed partially because of a missing slow pion or soft photon and lie in a broader range that still peaks around the signal region.

\par
The $\btag$ candidates are combined with $B$ mesons reconstructed in the decay $\Bdecay$; the latter are denoted as $\bsig$. The $\tau$ lepton is reconstructed as $\tautomu$, $e^+ \bar{\nu}_\tau \nu_e$, $\pi^+ \bar{\nu}_\tau$, and $\rho^+ \bar{\nu}_\tau$, with $\rho^+ \to \pi^+ \pi^0$.  
Since the neutrinos cannot be detected, the $\bsig$ candidate consists of a single charged track or a $\rho^+$ candidate. The $\rho^+$ candidate is required to have an invariant mass within $195 \mevcc$ of the nominal $\rho^+$ mass.
The signal-side decay modes are separated based on particle identification variables. The pion and kaon separation uses information from the ACC, TOF, and the $dE/dx$ measurement in the CDC; the electron identification is based on the same information in addition to the shape of the shower and the energy measurement in the ECL; and muon candidates are identified using hits in the KLM matched to a charged track. The selections of the signal-side decay modes are 
mutually exclusive. The momentum of the signal-side particle ($e^+, \mu^+, \pi^+$, or $\rho^+$) in the CM ($\psig$) must be in the range \mbox{$0.5 \gevc < \psig < 2.4 \gevc$.} 

\par
The combination of a $\btag$ and a $\bsig$ candidate provides an $\Upsilon (4S)$ candidate. The fact that the $\Upsilon (4S)$ is produced without any accompanying particles allows for a powerful selection: we reject events with additional $\pi^0$ candidates or charged tracks with $|dz| < 100$ cm and $|dr| < 20$ cm.
We also perform a selection in the remaining energy in the ECL ($\ecl$), defined as the sum of the energies of clusters in the ECL that are not associated with final state particles of the reconstructed $\Upsilon (4S)$ candidate. To mitigate beam-induced background in the energy sum, only clusters satisfying minimum energy thresholds of $50$, $100$, and $150 \mev$ are required for the barrel, forward, and backward end-cap calorimeter, respectively. Signal events peak near low values of $\ecl$ as only photons from beam-related background and misreconstructed events contribute, while the background is distributed over a much wider range. We require $\ecl$ to be smaller than $1.2 \gev$.

\par
In the decay channel $\tautoe$, a significant background arises from events containing converted photons. To suppress this, we combine the electron used in the reconstruction of either $\bsig$ or $\btag$ with every other oppositely-charged track in the event. Using the electron mass hypothesis for the second track, we require the invariant mass of the electron-track pair to be greater than $200 \mevcc$ for any of the pairs.
To suppress background from continuum events, we train another MVS with the following input variables: the angle of the $\btag$ candidate with respect to the beam direction in the CM; the angle between the thrust axis of the $\btag$ candidate and the remaining tracks in the event in the CM; 16 modified Fox-Wolfram moments~\cite{FWM}; and the momentum flow in nine concentric cones around the thrust axis of the $\btag$ candidate~\cite{CleoCones}. 
The background contributions differ significantly between the $\tau$ decay channels. Therefore, the requirements on the output of the two MVS (for the $\btag$ and continuum suppression) depend on the $\tau$ decay channel.
This is also true for the requirement on $\bdl$: it must be less than $1$ in all channels and greater than $-1.7$, $-1.9$, $-1.3$, and $-2.6$ for the $\tau$ decay channels to muon, electron, pion, and $\rho$, respectively.
The selections related to the output of the two MVS classifiers, $\bdl$, the particle identification of the $\pi^+$ from hadronic $\tau$ decay channels, and $M_{\pi^+ \pi^0}$ are optimized using samples of simulated signal and background events to give the highest figure-of-merit $N_S / \sqrt{N_S + N_B}$, where $N_S$ and $N_B$ are the number of expected signal and background events in the region $\ecl < 0.2 \gev$, respectively, as estimated from simulation and assuming ${\cal B}(\Bdecay) = 0.75 \times 10^{-4}$.

\par
The fraction of signal events having multiple candidates is $7\%$. In such events, we choose the candidate with the maximal value of the tag-side MVS classifier output. From MC simulation, we find that this method selects the correct candidate $70\%$ of the time.
The overall reconstruction and selection give a total reconstruction efficiency of $\epsilon = (23.1 \pm 0.1) \times 10^{-4}$, where the uncertainty is due to MC statistics only. It is described in detail in Table~\ref{tab:eff}. The given efficiencies contain all relevant branching fractions, including the corresponding branching fractions of the $\tau$ lepton.

\begin{table}[htb]
\caption{Reconstruction efficiency (in $10^{-4}$) for each $\tau$ decay mode, determined from MC and calibrated according to control sample studies. The uncertainty due to limited MC sample size is below $0.1$ for all values. The row denotes the generated decay mode, and the columns represent the reconstructed final state. The off-diagonal entries reflect the cross-feed between channels.}
\begin{tabular}{l c @{\hspace{1em}}c@{\hspace{1em}} c@{\hspace{1em}} c} 
\hline 
\hline 
        Final state & $e^+ \nu_e \bar{\nu}_\tau$         & $\mu^+ \nu_\mu \bar{\nu}_\tau$       & $\pi^+ \bar{\nu}_\tau$       & $\pi^+ \pi^0 \bar{\nu}_\tau$ \\ 
\hline 
$e^+ \nu_e \bar{\nu}_\tau$         & $6.6$         & $0.1$         & $0.2$         & $0.1$        \\ 
$\mu^+ \nu_\mu \bar{\nu}_\tau$     & $0.1$         & $4.7$         & $0.6$         & $0.2$        \\ 
$\pi^+ \bar{\nu}_\tau$             & $0$           & $0.1$         & $1.6$         & $0.5$        \\
$\pi^+ \pi^0 \bar{\nu}_\tau$       & $0$           & $0.1$         & $1.4$         & $4.9$        \\
$\pi^+ \pi^0 \pi^0 \bar{\nu}_\tau$ & $0$           & $0$           & $0.2$         & $1.3$        \\ 
Other                              & $0$           & $0$           & $0.1$         & $0.2$        \\ 
All                                & $6.8$         & $5.1$         & $4.0$         & $7.2$        \\
Total               & \multicolumn{4}{c}{$23.1$        } \\
                                        
\hline 
\hline 
\end{tabular}
\label{tab:eff}
\end{table}

\par
To study possible differences between real and simulated data, we use samples where $\bsig$ is reconstructed in the decays $B^+ \to \bar{D}^{*0} \ell^+ \nu_\ell$ and $B^+ \to \bar{D}^0 \pi^+$ (further denoted as double-tagged samples). The $D^{*0}$ mesons are reconstructed as $D^{*0} \to D^0 \pi^0$ and $D^0 \gamma$, and the $D^0$ meson as $D^0 \to K^- \pi^+$. The $D^{0}$ meson candidates are selected based on their mass. For the $D^{*0}$ meson candidates, additionally, the mass difference between the $D^{*0}$ and the $D^0$ meson candidates is used for selection. We apply the same four sets of different criteria on $\bdl$ and the two MVS classifiers corresponding to the four $\tau$ decay channels of the nominal analysis to each of the double-tagged $B$-decay samples. We measure the branching fractions of the $\bsig$ decays, determine the ratios with respect to the world average values~\cite{PDG}, and calculate weighted averages of the ratios of the $B$-decay channels. The reconstruction efficiency is found to be overestimated in MC simulation by a factor of $1.09 \pm 0.09, 1.08 \pm 0.08, 1.17 \pm 0.22$, and $1.02 \pm 0.10$ for the $\tau$ decay channels to muon, electron, pion, and $\rho$, respectively. The reconstruction efficiency is corrected based on this ratio, depending on the decay channel of the $\btag$ and the $\tau$.

\par
To extract the number of reconstructed signal events, we perform an extended two-dimensional unbinned maximum-likelihood fit in $\psig$ and $\ecl$.
We use smoothed histogram probability density functions (PDFs)~\cite{smooth} obtained from MC simulation to describe the signal and background components arising from events containing a $B\bar{B}$ pair. We use the product of one-dimensional PDFs for all components except for the signal in $\tautopi$ and $\tautorho$. In these modes, there is a significant cross-feed from channels with additional undetected neutral pions, resulting in a correlation between $\ecl$ and $\psig$ that is taken into account by using two-dimensional histogram PDFs. The continuum background, including $e^+ e^- \to q \bar{q} \ (q = u, d, s, c)$, $\tau^+ \tau^-$, and two-photon events, is described using the off-resonance data and is scaled according to the relative luminosities. Since the off-resonance data sample is very limited, its $\ecl$ distribution is described by a signal-mode-specific linear function. The slope of these functions is compatible with zero in all but the $\tautorho$ decay channel, which shows a slope of $37$ events per $\rm Ge\!V$. The uncertainties are $14, 6, 16$, and $14$ events per $\rm Ge\!V$, corresponding to a relative uncertainty of $20, 16, 25$, and $13\%$ on the number of events in the range $\ecl < 0.2 \gev$ for the $\tautomu, \tautoe, \tautopi$, and $\tautorho$ decay channel, respectively. The relative uncertainty on the normalization is about 10\% due to the limited size of the data sample.
\par
\begin{figure}[htb]
 \includegraphics[width=.40\textwidth]{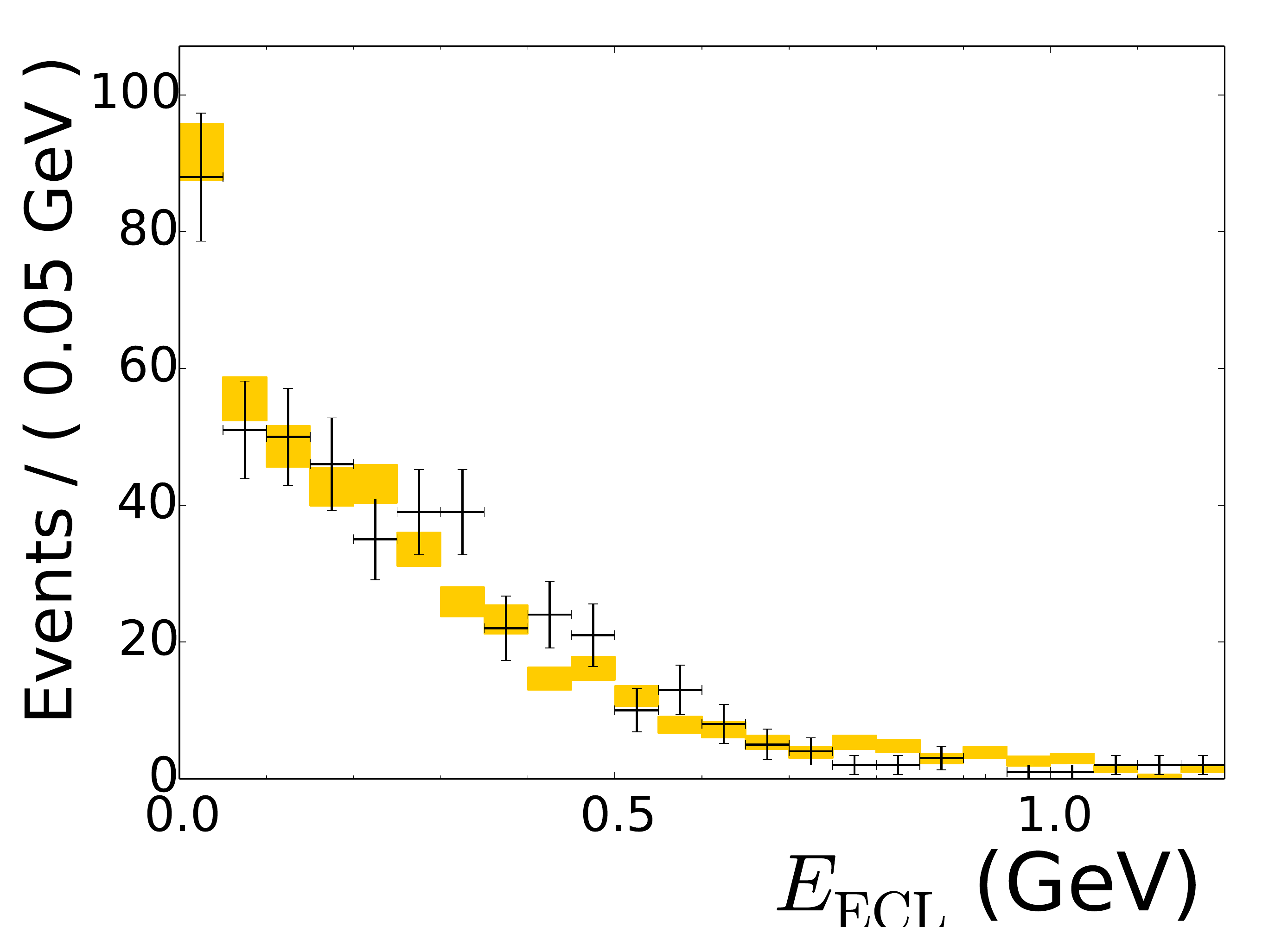} \includegraphics[width=.40\textwidth]{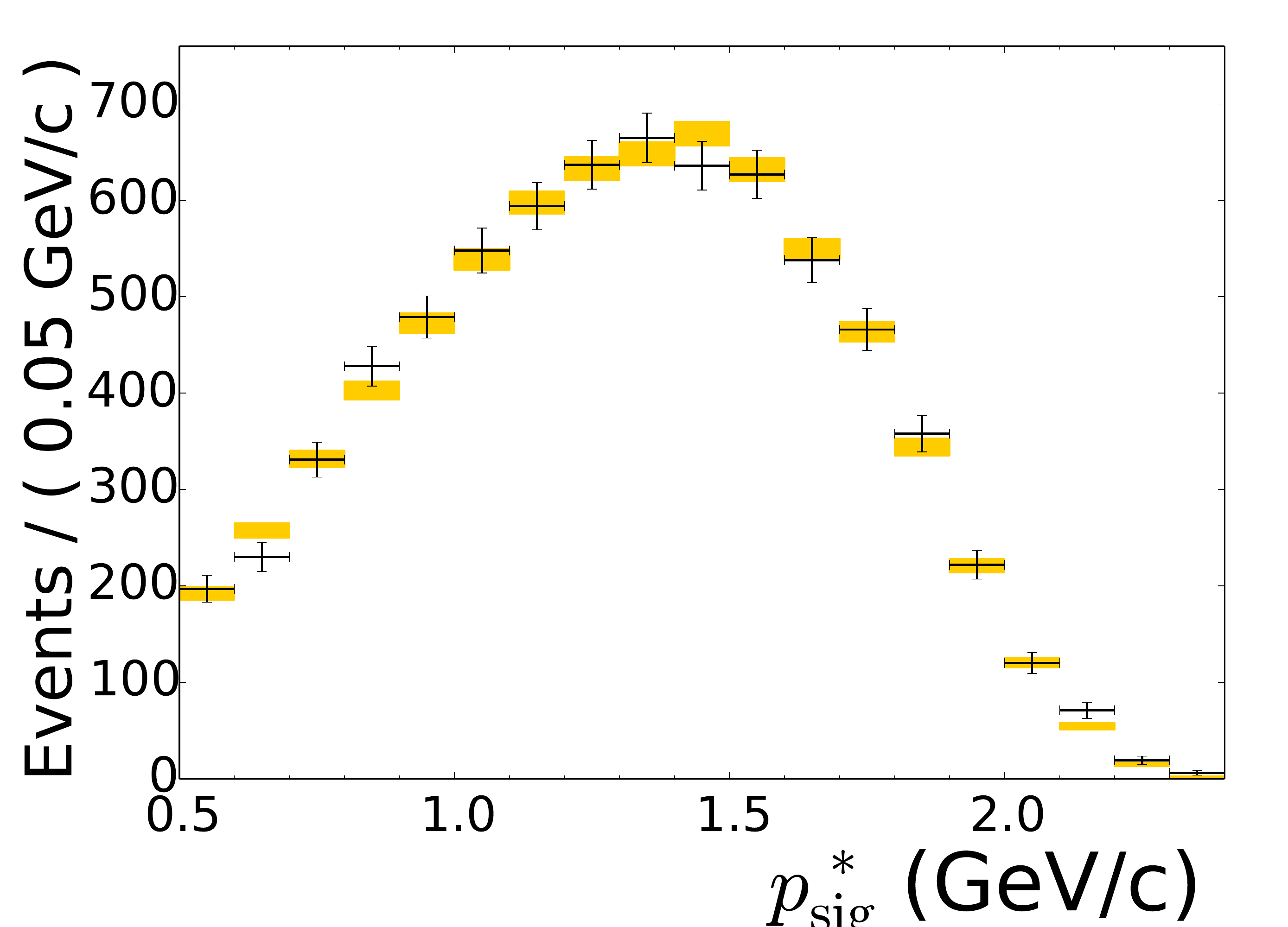}
 \caption{Comparison of the data and MC distribution in the double-tagged sample after the efficiency correction. The orange boxes show MC and the black markers the data. The upper and lower plots show the $\ecl$ distribution in the $B^+ \to \bar{D}^0 \pi^+$ sample and the $\psig$ distribution in the $B^+ \to \bar{D}^{*0} (\to \bar{D}^0 \pi^0) \ell^+ \nu_\ell$ sample, respectively. Both samples are selected corresponding to the $\tautomu$ decay channel.}
 \label{fig:comp}
\end{figure}

The ratio of the normalizations of the background components is fixed in the fit based on the yields estimated from simulation and off-resonance data. 
We compare the data and MC distribution of the signal component in $\ecl$ and $\psig$ using the double-tagged sample, which reveals no significant difference. This is illustrated in two representative plots in Fig.~\ref{fig:comp}. To validate the $\psig$ distributions, we treat the lepton from the $B$ decay as the $\tau$-decay product. We apply the same validation to other control samples and variables such as $\bdl$, the outputs of the MVS classifiers, and the missing energy in the event. 

\par
\begin{figure}[th]
 \includegraphics[width=.24\textwidth]{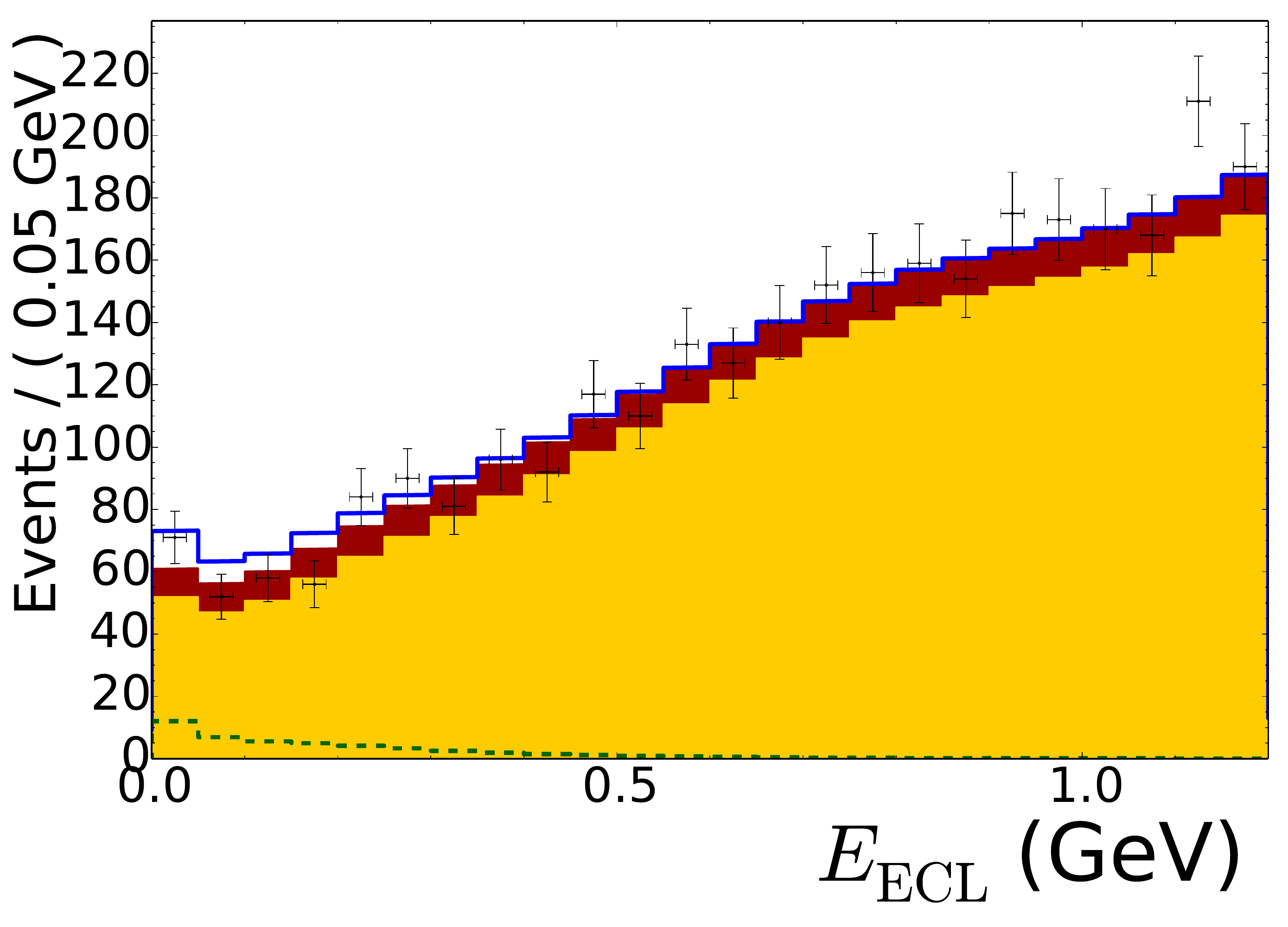} \put(-90,60){(a)} \includegraphics[width=.24\textwidth]{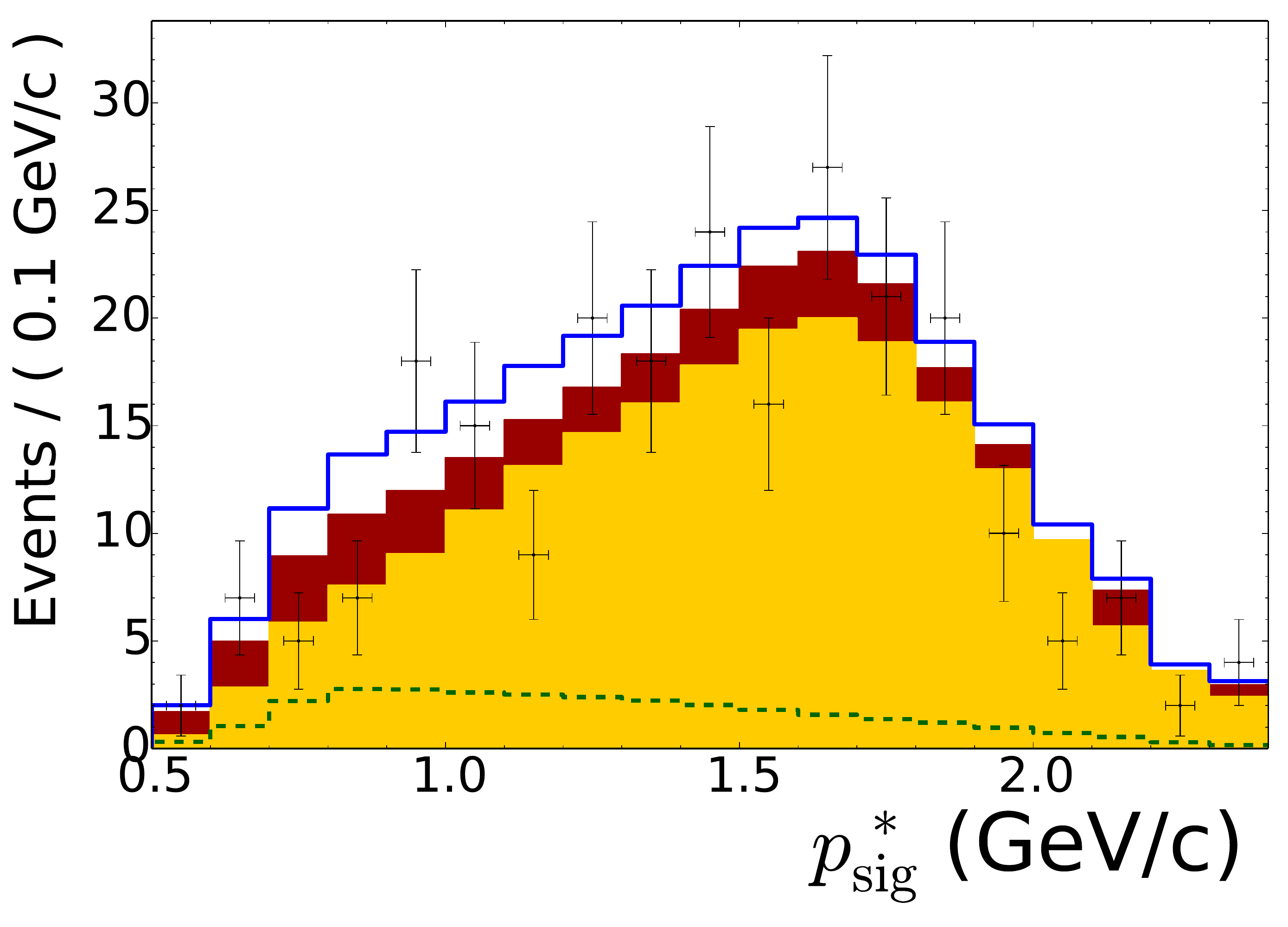} \put(-30,60){(a)} \\
 \includegraphics[width=.24\textwidth]{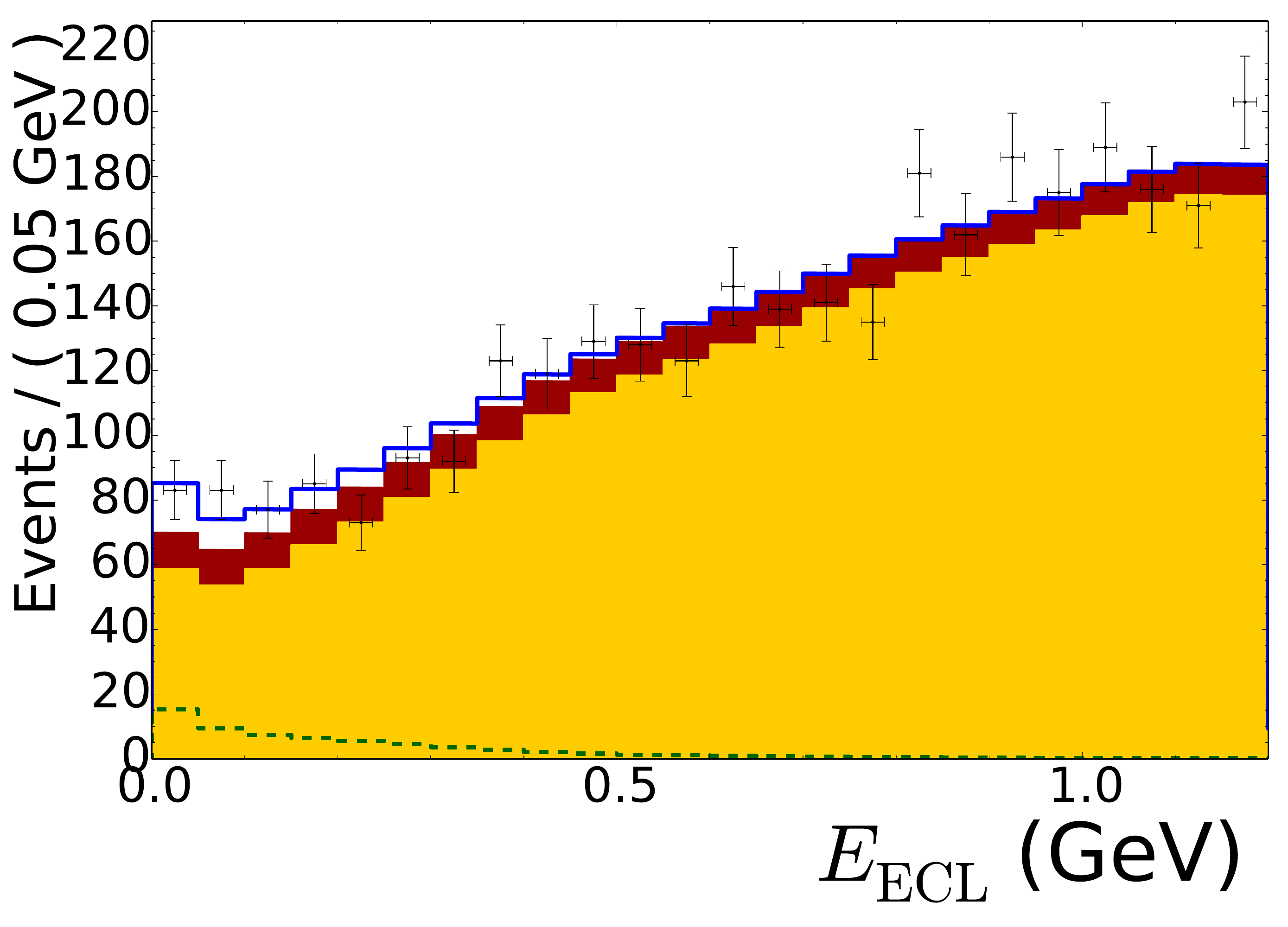} \put(-90,60){(b)} \includegraphics[width=.24\textwidth]{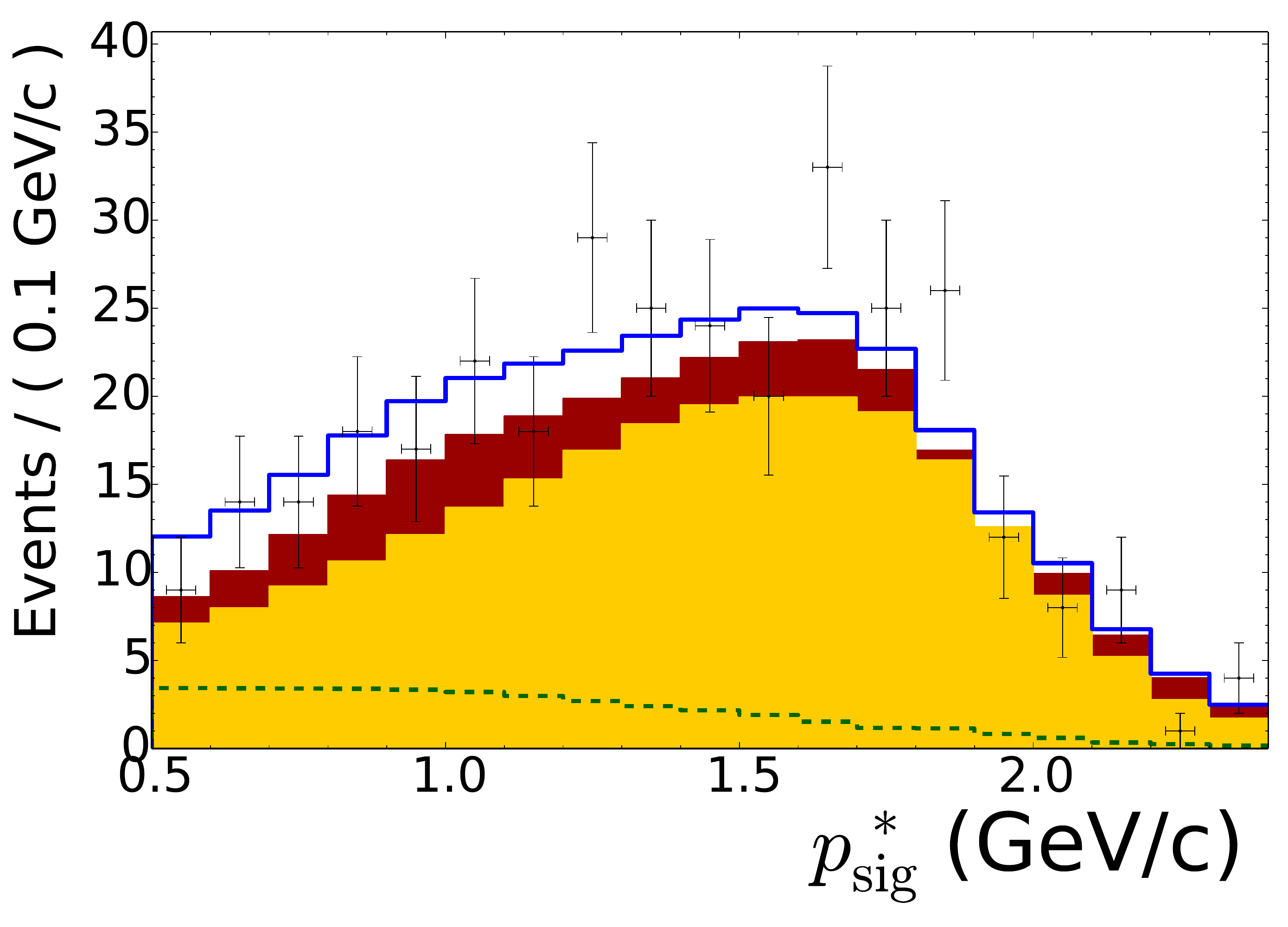} \put(-30,60){(b)} \\
 \includegraphics[width=.24\textwidth]{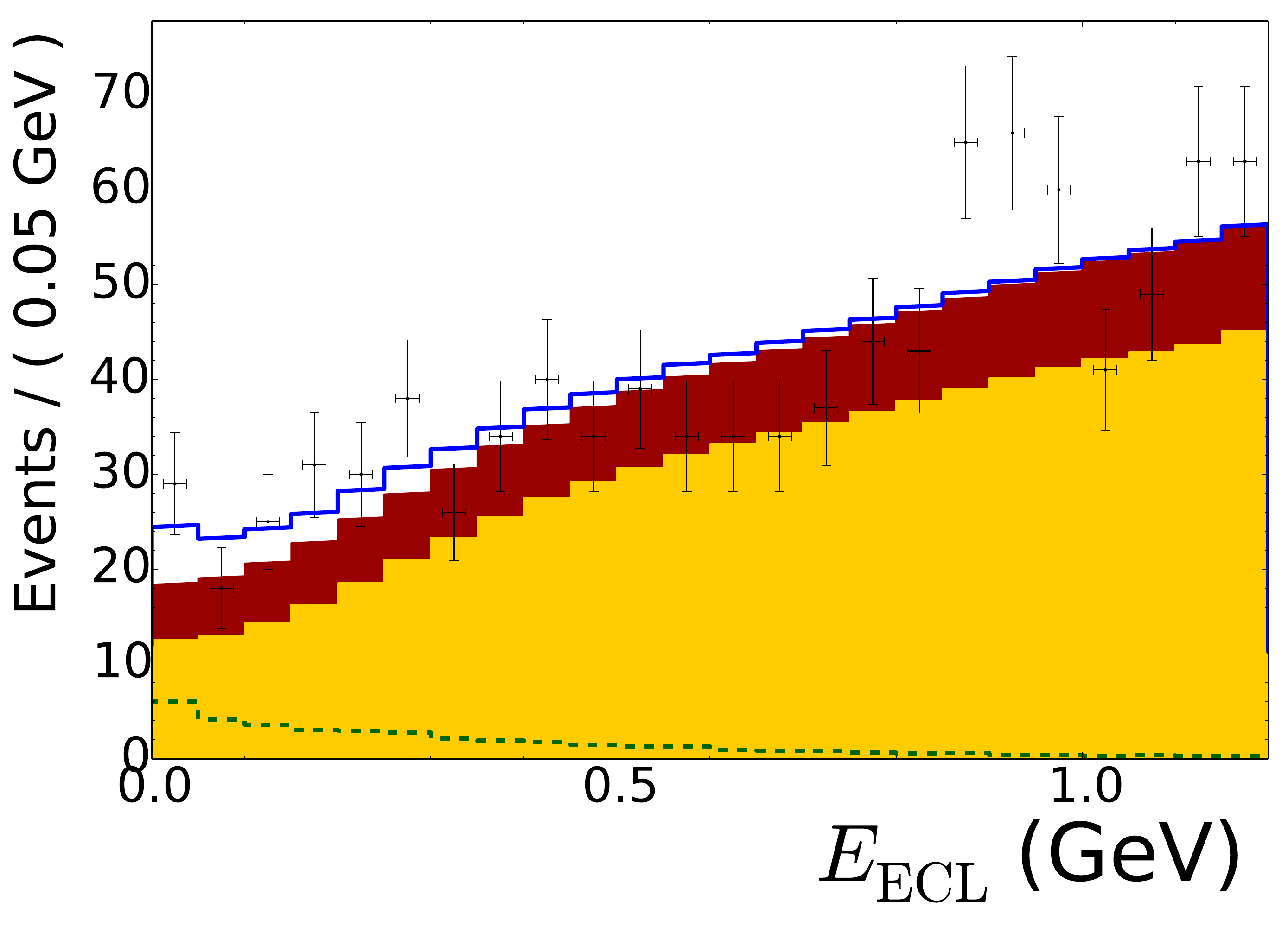} \put(-90,60){(c)} \includegraphics[width=.24\textwidth]{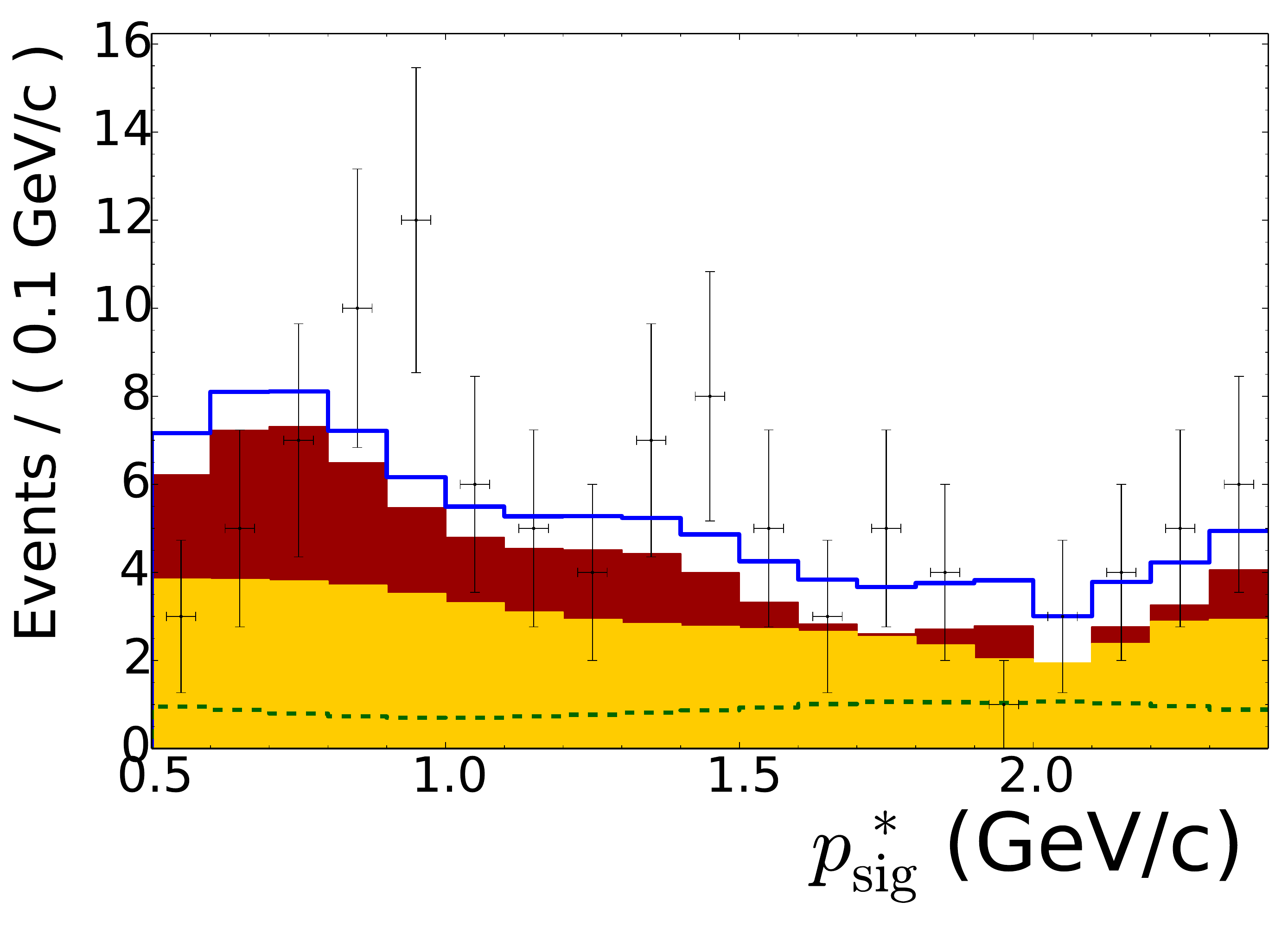} \put(-30,60){(c)} \\
 \includegraphics[width=.24\textwidth]{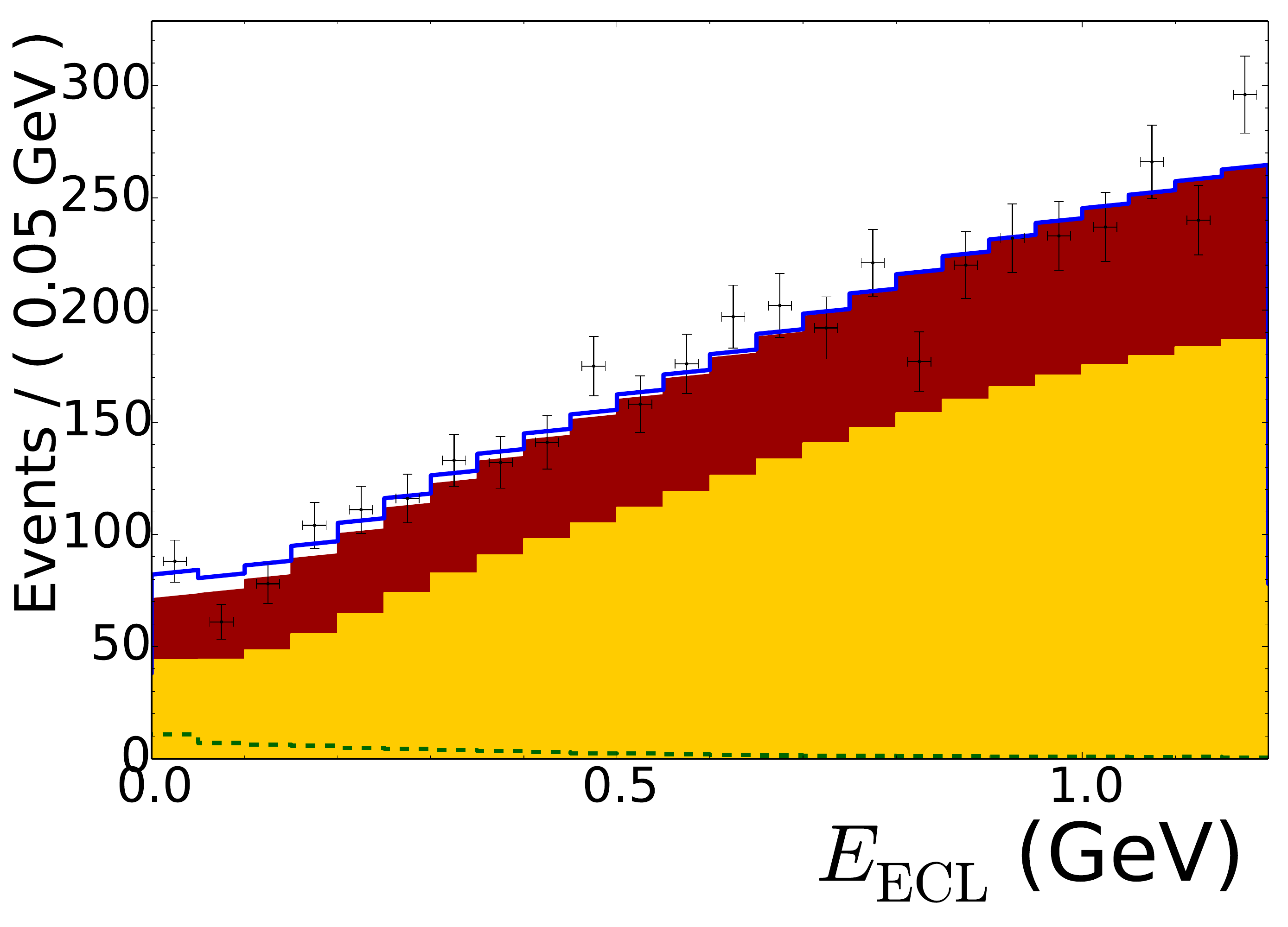} \put(-90,60){(d)} \includegraphics[width=.24\textwidth]{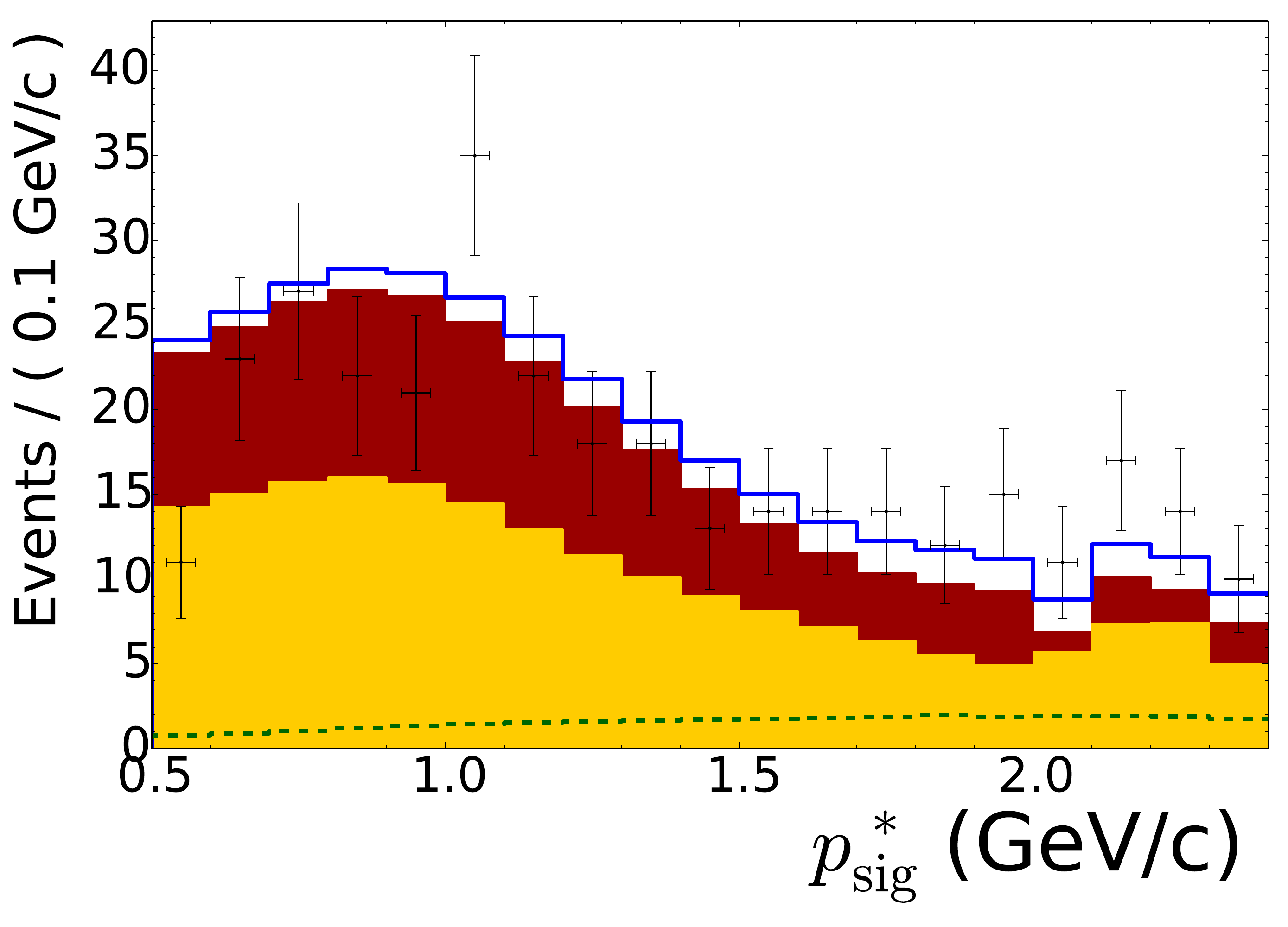} \put(-30,60){(d)}\\
 \includegraphics[width=.24\textwidth]{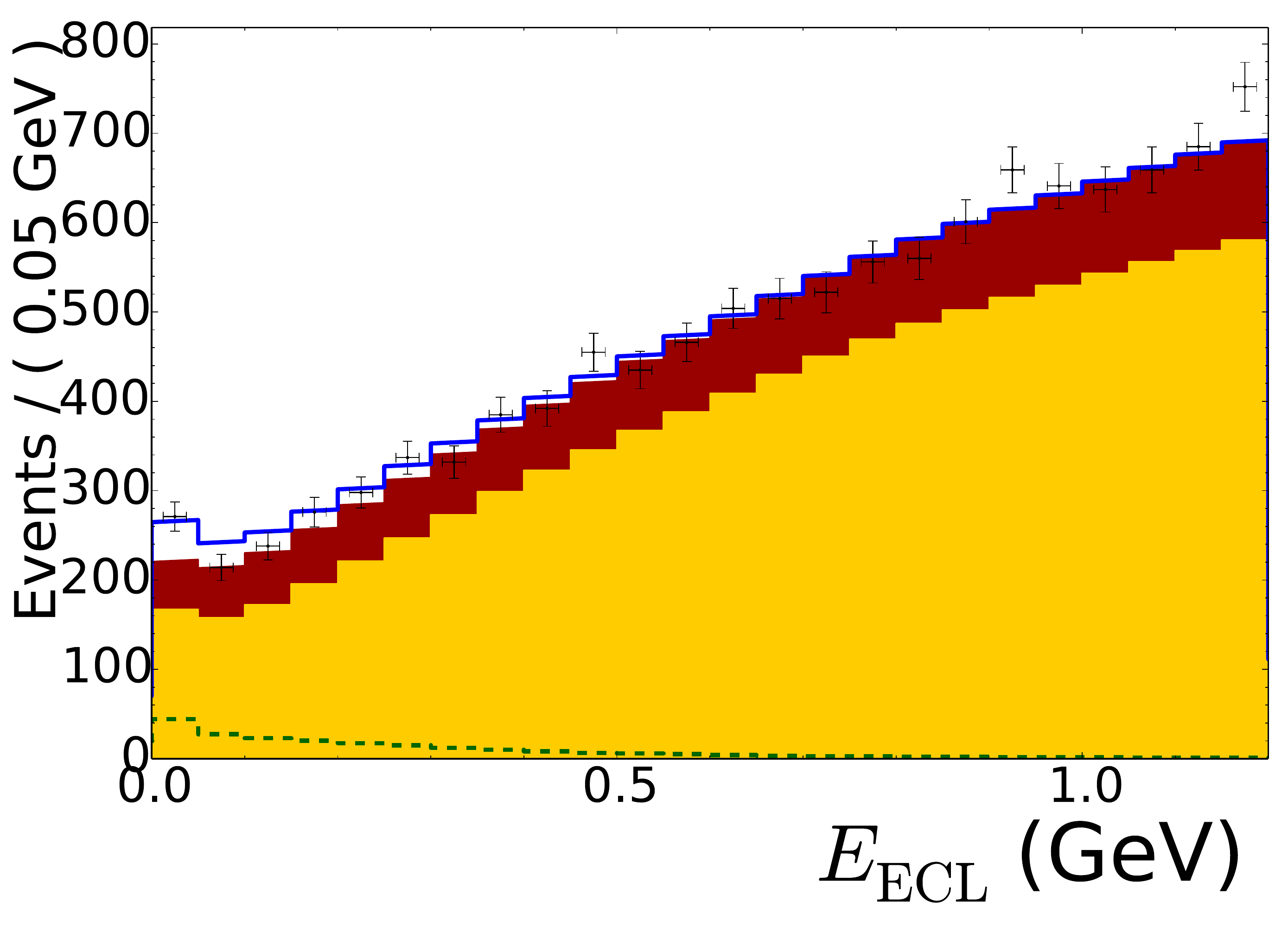} \put(-90,60){(e)} \includegraphics[width=.24\textwidth]{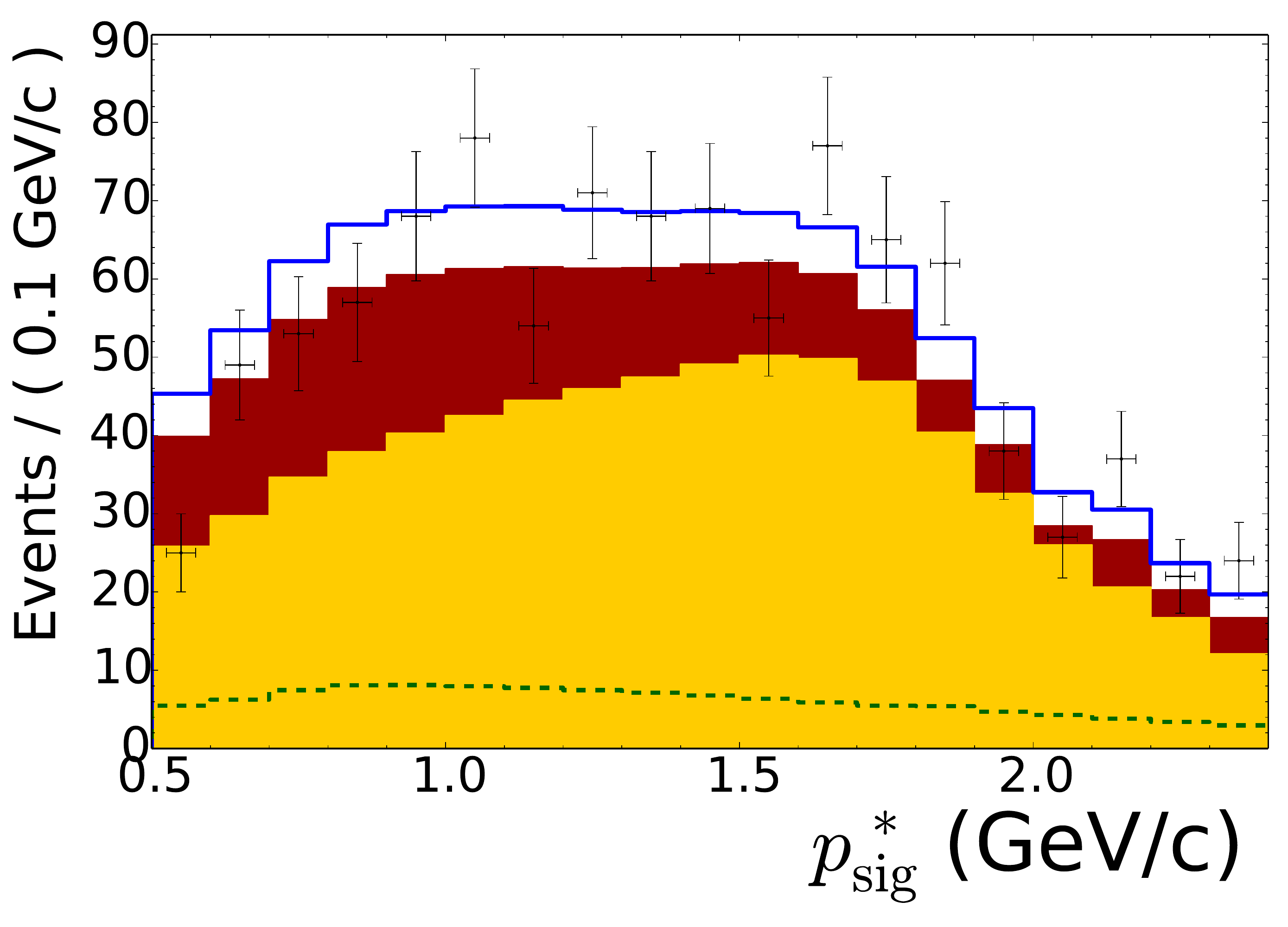} \put(-20,60){(e)}\\
\caption{Distributions for (a)~$\tautomu$, (b)~$\tautoe$, (c)~$\tautopi$, (d)~$\tautorho$, and (e) the sum of them. The left and right columns show the distributions of $\ecl$ and $\psig$ projected in the region $\ecl < 0.2$ GeV, respectively. The markers show the data distribution, the solid line the total fitted distribution, and the dashed line the signal component. The orange (red) filled distribution represents the $B \bar{B}$ (continuum) background.}
\label{Fig:plots}
 \end{figure}

The following five parameters vary in our final fit to the data: ${\cal B}(\Bdecay)$ and the normalization of the background in each of the four $\tau$ decay channels. The relative signal yields in the $\tau$ decay channels are fixed to the ratios of the reconstruction efficiencies. 
We obtain a total signal yield of $N_{\rm sig} = 222 \pm 50$, and this results in a branching fraction of ${\cal B}(\Bdecay) = (1.25 \pm 0.28) \times 10^{-4}$, where the uncertainties are statistical only. The signal yields and branching fractions, obtained from fits for each of the $\tau$ decay modes separately, are given in Table~\ref{tab:brSingle}. The results are consistent with a common value with a $p$ value of 10\%. The projections of the fitted distributions are shown in Fig.~\ref{Fig:plots}.

 \begin{table}[th]
 \caption{Signal yields and branching fractions, obtained from fits for the $\tau$ decay modes separately and combined. Errors are statistical only.}
  \begin{tabular}{l @{\hspace{1em}} r@{$\pm$}l @{\hspace{1em}} r@{$\pm$}l}
  \hline \hline
   Decay mode  & \multicolumn{2}{c}{$N_{\rm sig}$} & \multicolumn{2}{c}{${\cal B} (10^{-4})$} \\ \hline 
   $\tautomu$  & $13$  &$21$   & $0.34$  &$0.55$ \\
   $\tautoe$   & $47$  &$25$   & $0.90$  &$0.47$ \\
   $\tautopi$  & $57$  &$21$   & $1.82$  &$0.68$ \\
   $\tautorho$ & $119$ &$33$   & $2.16$  &$0.60$ \\
   Combined    & $222$ &$50$   & $1.25$  &$0.28$ \\
   \hline \hline
  \end{tabular}
  \label{tab:brSingle}
 \end{table}

\par 
The list of systematic uncertainties is given in Table~\ref{tab:sys}. The following systematic uncertainties are determined by varying the corresponding parameters by their uncertainty, repeating the fit and taking the difference to the nominal fit result as the systematic uncertainty: the normalization and slope of the continuum background component, where the dominant uncertainty originates from the error on the slope; the signal reconstruction efficiency; the branching fractions of the dominant background decays peaking in the $\ecl$ signal region, {\it e.g.}, $B^+ \to \bar{D}^0 \ell^+ \nu_\ell$ followed by $D^0 \to K_L K_L$ or $D^0 \to K_L K_L K_L$; the correction of the tagging efficiency, obtained from the double-tagged samples and assumed to be 100\% correlated among the four $\tau$ decay modes; and the branching fractions of the $\tau$ lepton. For branching fractions of $D$ mesons with multiple $K_L$ mesons in the final state, we use the values for corresponding decays with $K_S$ and take $50\%$ of the value as the uncertainty. 
\par
To estimate the effect of the uncertainty on the shape of the histogram PDFs due to the statistical uncertainty in the MC, the content of each bin is varied following a Poisson distribution with the initial value as the mean. This is repeated 1000 times and the standard deviation of the distribution of branching fractions is taken as systematic uncertainty. For the systematic uncertainty related to the best-candidate selection, we repeat the fit without applying this selection. The result is divided by the average multiplicity of $1.07$ and compared to the nominal fit result. The uncertainties on the efficiency of the reconstruction of charged tracks and neutral pions and on the efficiency of the particle identification have been estimated using high-statistics control samples. The charged-track veto is tested using the $D^0 \pi^+$ double-tagged sample by comparing the number of additional charged tracks in MC and data events. We find that it agrees well and so take the relative statistical uncertainty on the control sample as the systematic uncertainty. We also test an alternative description of the continuum background in $\ecl$ by using a polynomial of second order but the deviation is well covered by the related systematic uncertainty so we do not include it separately.
The quadratic sum of all contributions is $21.2 \%$. 

\begin{table}
\centering
\caption{List of systematic uncertainties.}
\begin{tabular}{l r}
\hline \hline
Source & Relative uncertainty (\%) \\ \hline
Continuum description 				& 14.1  \hspace{3em} ${}$\\
Signal reconstruction efficiency 		&  0.6  \hspace{3em} ${}$\\
Background branching fractions 			&  3.1  \hspace{3em} ${}$\\
Efficiency calibration			 	& 12.6  \hspace{3em} ${}$\\
$\tau$ decay branching fractions  		&  0.2  \hspace{3em} ${}$\\
Histogram PDF shapes  				&  8.5  \hspace{3em} ${}$\\
Best candidate selection			&  0.4  \hspace{3em} ${}$\\
Charged track reconstruction			&  0.4  \hspace{3em} ${}$\\
$\pi^0$ reconstruction 		 		&  1.1  \hspace{3em} ${}$\\
Particle identification 			&  0.5  \hspace{3em} ${}$\\
Charged track veto 				&  1.9  \hspace{3em} ${}$\\
Number of $B\bar{B}$ pairs			&  1.4  \hspace{3em} ${}$\\
Total & 21.2 \hspace{3em} ${}$\\
\hline \hline
\end{tabular}
\label{tab:sys}
\end{table}

\par
We find evidence for $\Bdecay$ decays with a significance of $3.8 \, \sigma$, by convolving the likelihood profile with a Gaussian whose width is equal to the systematic uncertainty. The significance is given by $\sqrt{2 \ln({\cal L} / {\cal L}_0) }$, where ${\cal L} ({\cal L}_0)$ is the value of the likelihood function when the signal yield is allowed to vary (set to 0).

\par
In summary, we report the measurement of the branching fraction of $\Bdecay$ decays using a sample of $772 \times 10^6$ $B\bar{B}$ pairs, which we analyze with the semileptonic tagging method. Our result is
\begin{equation*}
{\cal B}(\Bdecay) = [1.25 \pm 0.28 ({\rm stat.}) \pm 0.27({\rm syst.})] \times 10^{-4}
\end{equation*}
with a significance of $3.8 \ \sigma$. This result is consistent with our previous measurement based on the semileptonic tagging method of ${\cal B}(\Bdecay) = [1.54 \pm 0.38 ({\rm stat.}) \pm 0.37({\rm syst.})] \times 10^{-4}$~\cite{BelleLep} and supersedes it.
A combination with the recent Belle measurement based on the hadronic tagging method~\cite{BelleHad} of $[0.72^{+0.27}_{-0.25}({\rm stat.}) \pm 0.11 ({\rm syst.})] \times 10^{-4}$, taking into account all correlated systematic uncertainties, gives a branching fraction of ${\cal B}(\Bdecay) = [0.91 \pm 0.19 ({\rm stat.}) \pm 0.11({\rm syst.})] \times 10^{-4}$ with a combined significance of $4.6 \ \sigma$. This value is consistent with the SM expectation based on a fit using independent experimental input~\cite{CKMfitter}.

\par
We thank the KEKB group for the excellent operation of the 
accelerator; the KEK cryogenics group for the efficient
operation of the solenoid; and the KEK computer group,
the National Institute of Informatics, and the 
PNNL/EMSL computing group for valuable computing
and SINET4 network support.  We acknowledge support from
the Ministry of Education, Culture, Sports, Science, and
Technology (MEXT) of Japan, the Japan Society for the 
Promotion of Science (JSPS), and the Tau-Lepton Physics 
Research Center of Nagoya University; 
the Australian Research Council and the Australian 
Department of Industry, Innovation, Science and Research;
Austrian Science Fund under Grant No.~P 22742-N16 and P 26794-N20;
the National Natural Science Foundation of China under Contracts 
No.~10575109, No.~10775142, No.~10875115, No.~11175187, and  No.~11475187; 
the Ministry of Education, Youth and Sports of the Czech
Republic under Contract No.~LG14034;
the Carl Zeiss Foundation, the Deutsche Forschungsgemeinschaft
and the VolkswagenStiftung;
the Department of Science and Technology of India; 
the Istituto Nazionale di Fisica Nucleare of Italy; 
National Research Foundation (NRF) of Korea Grants
No.~2011-0029457, No.~2012-0008143, No.~2012R1A1A2008330, 
No.~2013R1A1A3007772, No.~2014R1A2A2A01005286, No.~2014R1A2A2A01002734, 
and No.~2014R1A1A2006456;
the Basic Research Lab program under NRF Grants No.~KRF-2011-0020333 and
No.~KRF-2011-0021196, Center for Korean J-PARC Users, No.~NRF-2013K1A3A7A06056592; 
the Brain Korea 21-Plus program and the Global Science Experimental Data 
Hub Center of the Korea Institute of Science and Technology Information;
the Polish Ministry of Science and Higher Education and 
the National Science Center;
the Ministry of Education and Science of the Russian Federation and
the Russian Foundation for Basic Research;
the Slovenian Research Agency;
the Basque Foundation for Science (IKERBASQUE) and 
the Euskal Herriko Unibertsitatea (UPV/EHU) under program UFI 11/55 (Spain);
the Swiss National Science Foundation; the National Science Council
and the Ministry of Education of Taiwan; and the U.S.\
Department of Energy and the National Science Foundation.
This work is supported by a Grant-in-Aid from MEXT for 
Science Research in a Priority Area (``New Development of 
Flavor Physics'') and from JSPS for Creative Scientific 
Research (``Evolution of Tau-lepton Physics'').


\begin{thebibliography}{99}
\bibitem{CC}
Throughout this paper, the inclusion of the charge-conjugate decay mode is implied unless otherwise stated.
\end{thebibliography}
\end{document}